\documentclass[english, aps, pra, twocolumn, superscriptaddress]{revtex4-1}
\usepackage[latin9]{inputenc}
\usepackage{color}
\usepackage{float}
\usepackage{amsmath}
\usepackage{amssymb}
\usepackage{graphicx}
\usepackage{mathptmx}
\allowdisplaybreaks[4]

\makeatletter
\usepackage[colorlinks,citecolor=blue,linkcolor=blue]{hyperref}
\usepackage{amsmath}

\makeatother

\usepackage{babel}
\begin{document}
\title{Coherence-Assisted Superradiant Laser with Hz Linewidth and $10^{-10}$W Power}
\author{Guohui Dong}
\affiliation{School of Physics and Electronic Engineering, Sichuan Normal University, Chengdu 610068, China}
\affiliation{Graduate School of Chinese Academy of Engineering Physics, Beijing 100084, China}
\author{Yao Yao}
\affiliation{Microsystem and Terahertz Research Center, China Academy of Engineering Physics, Chengdu 610200, China}
\author{Peng Zhang}
\email{pengzhang@ruc.edu.cn}

\affiliation{Department of Physics, Renmin University of China, Beijing, 100872,
China}
\affiliation{Beijing Computational Science Research Center, Beijing, 100084, China}
\author{Dazhi Xu}
\email{dzxu@bit.edu.cn}

\affiliation{Center for Quantum Technology Research and Key Laboratory of Advanced
Optoelectronic Quantum Architecture and Measurements (MOE), School
of Physics, Beijing Institute of Technology, Beijing 100081, China}
\affiliation{Beijing Computational Science Research Center, Beijing, 100084, China}
\begin{abstract}
The superradiant laser, based on the clock transition between the electric ground state $^1$S$_0$ and the metastable state $^3$P$_0$ of fermionic alkaline-earth(-like) atoms, has been proposed to be a new promising light source with linewidth being the order of millihertz. However, due to the small $^1$S$_0$-to-$^3$P$_0$ transition strength, the steady-state power in that system is  relatively low ($\sim 10^{-12}$W). In this work, we propose an alternative superradiant laser scheme based on a Raman-transition-induced coupling between the $^3$P$_0$ and $^3$P$_1$ states in bosonic alkaline-earth(-like) atoms, and achieve a laser with linewidth $\lesssim 2\pi\times1$Hz and power $\gtrsim 10^{-10}$W ($\sim 10^{3}$ photons in steady state) at a small pumping cost. The Raman beams play two significant roles in our scheme. First, the coherence between the dark and bright states induced by the Raman beams  produce a new local minimum in the pumping-linewidth curve with pumping rate lower than $2\pi \times 10$kHz, which is beneficial for continuous output. Second, the Raman beams mix the long-lived $^3$P$_0$ state into the lasing state and thus reduce the linewidth.  Our work greatly improves the output performance of the superradiant laser system with coherence induced by Raman transitions and may offer a firm foundation for its practical use in future.  
\end{abstract}
\maketitle

\section{Introduction}

Nowadays, the narrow-linewidth laser finds its importance in both fundamental scientific interest and practical applications, such as determining physical constants, gravitational wave detection, and global positioning systems~\citep{Abramovici1992,Fortier2007,Graham2013,Harry2006,Rosi2014}.
The traditional frequency stability method uses a rigid Fabry-Perot cavity as a reference, which is fundamentally limited by thermal fluctuations of the cavity length~\citep{Numata2004,Notcutt2005,Notcutt2006,Kessler2012,Cole2013}.
In the past decade, instead of further reduction of thermal
noise of the reference cavity, the superradiant laser that utilizes
the atomic transition with a long lifetime, such as the $^1\mathrm{S}_0$-to-$^3\mathrm{P}_0$ clock transition
of fermionic alkaline-earth atoms, has been proposed and attracted much attention. Unlike the traditional laser, superradiant laser works in the bad-cavity regime where the cavity loss is several orders of magnitude larger than the atomic decay~\citep{Chen2009,Meiser2009,Bohnet2012,Norcia2016a,Norcia2016,Holland2017,Debnath2018,Norcia2018a}. Thus the cavity mode can be eliminated adiabatically in this regime, leading to a strong atom-atom correlation (a collective spin)~\citep{Meiser2010,Bohnet2012,Liu2020,Shankar2021}, and the coherence of the superradiant laser is solely stored in the atoms. The most attractive feature of the superradiant laser is that its linewidth can be smaller than the corresponding atomic decay rate with frequency robust against cavity length fluctuation~\citep{Meiser2009,Norcia2016}.

However, in previous superradiant laser systems, it is hard to achieve a laser with both narrow linewidth and large power at the same time. For instance, due to the long-lifetime nature of the $^3\mathrm{P}_0$ state in fermionic alkaline-earth atoms, the weak coupling of the $^1\mathrm{S}_0$-to-$^3\mathrm{P}_0$ transition to the cavity field prevents the system from outputting a large power, while the scheme with bosonic alkaline-earth atoms possesses a stronger coupling to the cavity field but a much broader laser linewidth for about $2\pi\times$kHz~\citep{Norcia2016,Debnath2018}. Considering that the superradiant laser has been experimentally demonstrated based on the $^1\mathrm{S}_0$-to-$^3\mathrm{P}_0$ transition in $^{87}$Sr~\citep{Norcia2016a} and the $^1\mathrm{S}_0$-to-$^3\mathrm{P}_1$ transition in $^{88}$Sr~\citep{Norcia2016}, we expect to propose a new scheme with both the advantages of these two systems, i.e., narrow linewidth and appreciable power.

To this end, we propose to couple the $^3$P$_0$ state to the $^3$P$_1$ state in the bosonic alkaline-earth atom (e.g., $^{88}$Sr) by Raman transitions. The $^3$P$_0$ state has an extremely long lifetime which can effectively suppress the spontaneous decay rate of the lasing state. Meanwhile, the laser power can get significantly increased compared to the system with fermionic alkaline-earth atoms, since the $^1\mathrm{S}_0$-to-$^3$P$_1$ transition assists in the lasing process. The idea of this proposal stems from our previous work~\citep{Dong2020} where we produce a Raman-transition-assisted ultra-narrow transmission spectrum based on magnetically induced optical transparency~\citep{Winchester2017}.

The second-order mean-field calculation in $^{88}$Sr atomic ensemble shows that the laser linewidth has two local minimums with the change of the pumping rate. For proper Raman strengths and ratios, the left minimum of linewidth becomes the global minimum ($< 2\pi\times1$Hz) with the pumping rate as low as about $2\pi\times10$kHz. Hence, our scheme can realize a superradiant laser with $2\pi \times 1$Hz-level linewidth and $10^{-10}$W-level power ($\sim 10^{3}$ photons in steady state) at a small pump.  The feature of this narrow linewidth at a low pumping rate can inhibit the heating effect in the lasing process and thus is very beneficial for continuous output. Further analyses and calculations demonstrate that this double-minimum character in linewidth originates from the coherence between the dark and bright states induced by the Raman transitions.

\section{System Setup and Lasing Process\label{sec:model}}

\subsection{\label{IIA}System Setup}

\begin{figure}[t]
\begin{centering}
\includegraphics[scale=0.6]{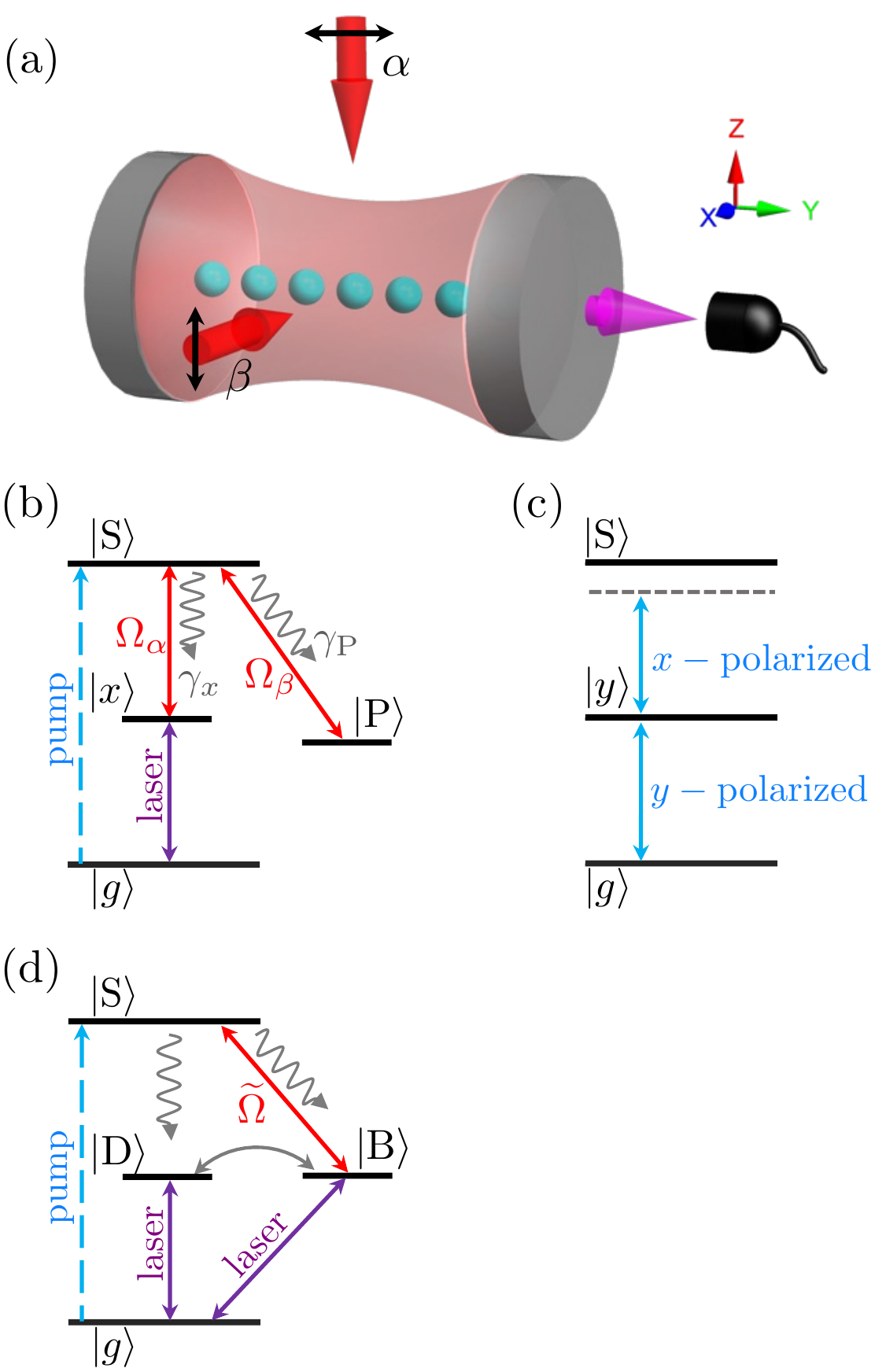}
\par\end{centering}
\caption{\label{fig1}{\bf (a)}: $N$ bosonic alkaline-earth(-like) atoms are trapped in an optical cavity. Two Raman beams $\alpha$ and $\beta$ polarized along the $y$- and $z$-directions, respectively, are injected into the cavity. {\bf (b)}:
Schematic diagram of the four-level lasing system.
The atoms are incoherently pumped to the $^3$S$_1$ state $|{\rm S}\rangle$ and then transit to the $^{3}\mathrm{P}_{1}$ state $\vert x\rangle$ or the $^{3}\mathrm{P}_{0}$ state $\vert {\rm P}\rangle$ via spontaneous decay or Raman-induced coupling.
The atoms in $\vert x\rangle$ further transit to the $^1$S$_0$ state $\vert g\rangle$
and emit laser photons into the cavity mode.
{\bf (c)}: A possible scheme of pumping the atoms to state $|{\rm S}\rangle$ with two beams polarized along the $x$- and $y$-directions. {\bf (d)}:
Schematic diagram of our scheme in the basis with dark state $|{\rm D}\rangle$ and bright state $|{\rm B}\rangle$. The atoms in state $|{\rm S}\rangle$
can transit to $|{\rm D}\rangle$ via only spontaneous decay and to $|{\rm B}\rangle$ via both spontaneous decay and the Raman transition. The atoms in both $|{\rm B}\rangle$ and $|{\rm D}\rangle$ can emit laser photons.}
\end{figure}

In our system, as shown in Figs.~\ref{fig1}(a) and (b), an ensemble of $N$ cold bosonic alkaline-earth(-like) atoms
are trapped in an optical cavity with the axis along the $y$-direction. The quantization axis of the atomic angular momentum is along the $z$-direction, and $J_z=m_{J} \hbar$ (we set $\hbar=1$ hereinafter) is the projection on the $z$-axis. The $x$-polarized cavity mode couples with the lasing transition between the electronic $^{1}\mathrm{S}_{0}$ state $\vert g\rangle$ and the $^{3}\mathrm{P}_{1}$ state $|x\rangle$ which is defined as
\begin{eqnarray}
\vert x\rangle\equiv\frac{1}{\sqrt{2}}(|^{3}\mathrm{P}_{1},m_J=-1\rangle-|^{3}\mathrm{P}_{1},m_J=+1\rangle ).
\end{eqnarray}
Two Raman beams, $\alpha$ and $\beta$, are employed to effectively couple the $^{3}\mathrm{P}_{1}$ state to the long-lived $^{3}\mathrm{P}_{0}$ state. The $y$-polarized beam $\alpha$ couples the state $|x\rangle$ to the $^{3}\mathrm{S}_{1}$ ($m_J=0$) state $|\rm S\rangle$, and the $z$-polarized beam $\beta$ couples $|{\rm S}\rangle$ to the $^{3}\mathrm{P}_{0}$ state $|{\rm P}\rangle$.

\subsection{\label{lm}Lasing Process}

Before the detailed calculation, it is necessary to present a rough and qualitative description of how the laser is generated in our scheme. The pumping process incoherently transfers the atoms from the ground state $|g\rangle$ to the excited state $|{\rm S}\rangle$, which can be accomplished in two steps as shown in Fig.~\ref{fig1}(c): first, a $y$-polarized beam pumps the atoms from $|g\rangle$ to the $^3$P$_1$ state $\vert y\rangle\equiv i(\vert ^{3}\mathrm{P}_{1},m_J=-1\rangle+\vert ^{3}\mathrm{P}_{1},m_J=+1\rangle)/\sqrt{2}$; then, an $x$-polarized beam pumps the atoms from $\vert y\rangle$ to $|{\rm S}\rangle$. As long as these two driving beams are largely detuned with the electronic transitions, the pumping process can be considered incoherent.

Part of the atoms in the $^3$S$_1$ state will fall to the state $|x\rangle$ or $|{\rm P}\rangle$ via either the spontaneous decay or Raman-beam-induced coherent transitions. As the $^{3}\mathrm{P}_{0}$-to-$^{1}\mathrm{S}_{0}$ transition of bosonic alkaline-earth atoms is electric-dipole forbidden (e.g., $^{24}$Mg, $^{40}$Ca, $^{88}$Sr)~\citep{Santra2004}, it does not couple with the cavity mode. Hence, when the population inversion is established between $|x\rangle$ and $|g\rangle$ and the stimulated emission overwhelms the photon loss, the laser polarized in the $x$-direction will be generated. 

Although the $^{1}\mathrm{S}_{0}$-to-$^{3}\mathrm{P}_{0}$ transition does not directly couple with the laser mode, the Raman-induced coherence between the $^{3}\mathrm{P}_{0}$ and $^{3}\mathrm{P}_{1}$ states plays a crucial role in reducing the linewidth, which is similar to our previous work on the ultra-narrow transmitted spectrum~\citep{Dong2020}. An appropriate way to investigate this coherence is revisiting the above lasing scheme utilizing the dark ($|{\rm D}\rangle$) and bright ($|{\rm B}\rangle$) states corresponding to the Raman coupling, which are defined as:
 \begin{eqnarray}
\left|{\rm D}\right\rangle &\equiv&
(\Omega_{\beta}\vert x\rangle-\Omega_{\alpha}\vert\text{P}\rangle)/\widetilde{\Omega},\label{dark} \\
\left|{\rm B}\right\rangle &\equiv&
(\Omega_{\alpha}\vert x\rangle+\Omega_{\beta}\vert\text{P}\rangle)/\widetilde{\Omega}.\label{bright}
\end{eqnarray}
Here, $\Omega_{\alpha(\beta)}$ is the Rabi frequency corresponding to the Raman beam $\alpha\;(\beta)$, and $\widetilde{\Omega}\equiv\sqrt{\vert\Omega_{\alpha}\vert^{2}+\vert\Omega_{\beta}\vert^{2}}$ is the Raman strength.

Figure~\ref{fig1}(d) shows the same lasing scheme as  Fig.~\ref{fig1}(b) but uses the basis of $|{\rm D}\rangle$ and $|{\rm B}\rangle$. We can see that the state $|{\rm S}\rangle$ only couples with the bright state $|{\rm B}\rangle$ through the Raman beams with effective coupling intensity $\widetilde{\Omega}$. Thus besides the spontaneous decay, the pumped state $|{\rm S}\rangle$ can also transit to  the state $|{\rm B}\rangle$ via the Raman transition. In contrast, the state $|{\rm S}\rangle$ transits to the state $|{\rm D}\rangle$ only through the spontaneous decay. As both the states $|{\rm D}\rangle$ and $|{\rm B}\rangle$ have the $|x\rangle$ component, the lasing mode coherently couples with both the $|{\rm D}\rangle$-$|g\rangle$ and $|{\rm B}\rangle$-$|g\rangle$ transitions.  It is remarkable that there exists an incoherent coupling between $|{\rm D}\rangle$ and $|{\rm B}\rangle$, which comes from the spontaneous decay of the state $|{\rm S}\rangle$ \footnote{One can find this coupling by writing the master equation Eq.~(\ref{eq:master}) in the basis of $|{\rm D}\rangle$ and $|{\rm B}\rangle$}. We will show the explicit calculation of the laser properties using the bare basis $|x\rangle$ and $|{\rm P}\rangle$ in Sec.~\ref{sec:mean-field} and Appendix~\ref{AppA}, and discuss the influence of coherence between the dark and bright lasing transitions on the laser linewidth in Sec.~\ref{sec:V}.

\section{Model and methods \label{sec:mean-field}}

In this section, we use the second-order mean-field master equation to calculate the laser power, frequency and linewidth in our scheme.

\subsection{Hamiltonian and Master Equation}

The Hamiltonian, including the $N$ atoms, the quantized cavity field, and the classical Raman beams, is given by
\begin{eqnarray}
&&\hat{H} =\omega_{c}\hat{a}^{\dagger}\hat{a}+\sum_{j=1}^{N}\left(\omega_{0}\hat{\sigma}_{xx}^{(j)}+\omega_{\text{S}}\hat{\sigma}_{\text{SS}}^{(j)}+\omega_{\text{P}}\hat{\sigma}_{\text{PP}}^{(j)}\right)\nonumber \\
 &&+\sum_{j=1}^{N}\left(\frac{\Omega_{c}}{2}\hat{a}^{\dagger}\hat{\sigma}_{gx}^{(j)}+\frac{\Omega_{\alpha}}{2}\hat{\sigma}_{\text{S}x}^{(j)}e^{-i\omega_{\alpha}t}+\frac{\Omega_{\beta}}{2}\hat{\sigma}_{\text{SP}}^{(j)}e^{-i\omega_{\beta}t}+\rm{h.c.}\right),\nonumber\\
 \label{eq:H_t}
\end{eqnarray}
where $\hat{a}~(\hat{a}^{\dagger})$ is the annihilation (creation) operator
of the cavity mode with angular frequency $\omega_{c}$ and
\begin{eqnarray}
\hat{\sigma}_{\mu\nu}^{(j)}\equiv\vert\mu\rangle^{(j)}\langle\nu\vert\ \ (\mu,\nu=g,x,\text{S},\text{P})
\end{eqnarray}
is the electronic state transition operator for the $j$th atom. We choose the energy of the electronic ground state $|g\rangle$ to be zero, and then the frequencies of the states $|x\rangle$, $|{\rm S}\rangle$ and $|{\rm P}\rangle$ are $\omega_{0}$, $\omega_{\text{S}}$ and $\omega_{\text{P}}$, respectively. The angular frequencies of the Raman beams $\alpha$ and $\beta$ are $\omega_\alpha$ and $\omega_\beta$, respectively. $\Omega_{c}$ is the Rabi frequency of the atom-cavity coupling, and $\Omega_{\alpha (\beta)}$ is that of the Raman-induced coupling between the state $|{\rm S}\rangle$ and $|x\rangle$ $(|\rm P\rangle)$. We assume that all the atoms homogeneously interact with the optical cavity and the Raman beams. Without loss of generality, we choose all the Rabi frequencies as real.

In the rotated frame, the above Hamiltonian Eq.~(\ref{eq:H_t}) can be simplified as
\begin{eqnarray}
\hat{H}_{I} & =&\delta_{c}\hat{a}^{\dagger}\hat{a}+\sum_{j=1}^{N}\left[\delta_{\alpha}\hat{\sigma}_{\text{SS}}^{(j)}+(\delta_{\alpha}-\delta_{\beta})\hat{\sigma}_{\text{PP}}^{(j)}\right]\nonumber \\
 && +\sum_{j=1}^{N}\left(\frac{\Omega_{c}}{2}\hat{a}^{\dagger}\hat{\sigma}_{gx}^{(j)}+\frac{\Omega_{\alpha}}{2}\hat{\sigma}_{\text{S}x}^{(j)}+\frac{\Omega_{\beta}}{2}\hat{\sigma}_{\text{SP}}^{(j)}+\rm{h.c.}\right),\label{eq:H_I}
\end{eqnarray}
where the detunings are defined as
\begin{eqnarray}
\delta_{c} &\equiv&\omega_{c}-\omega_{0},\label{dc}\\
\delta_{\alpha} &\equiv&\omega_{\text{S}}-\omega_{0}-\omega_{\alpha},\label{da}\\
\delta_{\beta} &\equiv&\omega_{\text{S}}-\omega_{\text{P}}-\omega_{\beta}.\label{db}
\end{eqnarray}
In the following, when we explore the laser power and linewidth, all these detunings are assumed to be zero. When calculating the pulling coefficients which describe the influence of these detunings on the laser frequency, we assume that they fluctuate around zero.

During the incoherent pumping process, the spontaneous decay of the electronic excited states and the loss of the photon from the cavity are also taken into account. The  evolution of the atoms and cavity field density matrix $\hat{\rho}(t)$ is determined by the Born-Markov master equation
\begin{eqnarray}
\frac{d}{dt}\hat{\rho} &=&-i[\hat{H}_{I},\hat{\rho}]+\kappa\mathcal{L}[\hat{a}]\hat{\rho}\nonumber \\
&& +\sum_{j=1}^{N}\left(\gamma_{x}\mathcal{L}[\hat{\sigma}_{x\text{S}}^{(j)}]+\gamma_{\text{P}}\mathcal{L}[\hat{\sigma}_{\text{PS}}^{(j)}]\right)\hat{\rho}\nonumber \\ &&+\sum_{j=1}^{N}\left(\gamma_{0}\mathcal{L}[\hat{\sigma}_{gx}^{(j)}]+\eta\mathcal{L}[\hat{\sigma}_{\text{S}g}^{(j)}]\right)\hat{\rho}, \label{eq:master}
\end{eqnarray}
with the Lindblad operator defined as
\begin{equation}
\mathcal{L}[\hat{\mathcal{O}}]\hat{\rho}\equiv\hat{\mathcal{O}}\hat{\rho}\hat{\mathcal{O}}^{\dagger}-\frac12\left(\hat{\mathcal{O}}^{\dagger}\hat{\mathcal{O}}\hat{\rho}+\hat{\rho}\hat{\mathcal{O}}^{\dagger}\hat{\mathcal{O}}\right).
\end{equation}
Here, $\kappa$ is the decay rate of the cavity photon, and  $\gamma_{0}$ ($\gamma_{x}$,  $\gamma_{\rm P}$) is the spontaneous decay rate of the state $\vert x\rangle$ to $|g\rangle$ ($|\rm S\rangle$ to $| x\rangle$, $|\rm S\rangle$ to $|\rm P\rangle$). $\eta$ characterizes the effective pumping rate from the ground state $|g\rangle$ to the state $|\rm S\rangle$.

\subsection{Calculation of Laser Power \label{clp}}

\begin{figure*}[t]
\begin{centering}
\includegraphics[scale=0.24]{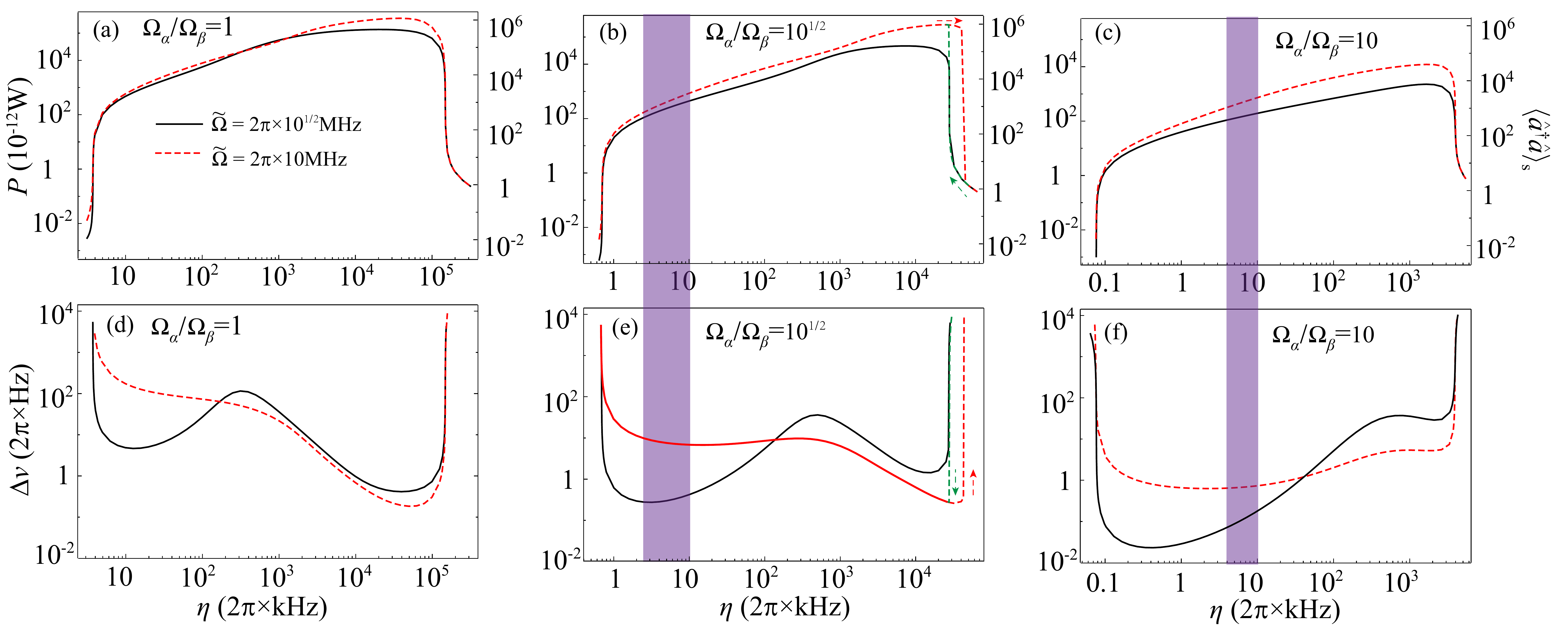}
\par\end{centering}
\caption{\label{fig2}The output power (average photon number)  and linewidth of the superradiant laser as functions of pumping rate for the cases with $\Omega_{\alpha}/\Omega_{\beta}=1$ {\bf (a, d)}, $10^{1/2}$ {\bf (b, e)}, and $10$ {\bf (c, f)}. The black-solid and red-dashed lines are plotted for $\widetilde{\Omega}=2\pi\times10^{1/2}$MHz and $2\pi\times10$MHz, respectively. The regions satisfying $P\protect\geq 10^{-10}$W,
$\Delta\nu\protect\leq2\pi\times1$Hz, and $\eta\protect\leq2\pi\times10$kHz can be found when $\Omega_{\alpha}/\Omega_{\beta}=10^{1/2}$ and $10$, as marked in purple. In {\bf(b)}, optical bistability appears when $\widetilde{\Omega}=2\pi\times10$MHz and the pumping strength is around the greater laser threshold ($\eta\sim 2\pi\times20$MHz), where the red (green) dashed line plots the laser power in the direction of increasing (decreasing) the pumping rate.}
\end{figure*}

In this work, we solve the master equation Eq.~(\ref{eq:master}) with the second-order
mean-field approximation, where the correlations of three and more operators are ignored during the cumulant expansion~\citep{Meiser2009,Zhang2021,Kubo1962}, for example,
\begin{eqnarray}
\langle\hat{A}\hat{B}\hat{C}\rangle\simeq\langle\hat{A}\rangle\langle\hat{B}\hat{C}\rangle+\langle\hat{B}\rangle\langle\hat{A}\hat{C}\rangle+\langle\hat{C}\rangle\langle\hat{A}\hat{B}\rangle-2\langle\hat{A}\rangle\langle\hat{B}\rangle\langle\hat{C}\rangle.\nonumber\\
\end{eqnarray}
Here, we define the instantaneous and steady-state expectation values of operator $\hat{\mathcal{O}}$ as
\begin{eqnarray}
\langle\hat{\mathcal{O}}\rangle&\equiv&\text{Tr}[\hat{\mathcal{O}}\hat{\rho}(t)],\\
\langle\hat{\mathcal{O}}\rangle_{\text{s}}&\equiv&\text{Tr}[\hat{\mathcal{O}}\hat{\rho}(\infty)].
\end{eqnarray}

As all the coupling strengths are homogeneous with respect to the $N$ atoms, the one- and two-operator expectation values are symmetric concerning the permutation of atoms. Therefore, we can write
$\langle\hat{\sigma}_{\mu\nu}^{(j)}\rangle=\langle\hat{\sigma}_{\mu\nu}^{(1)}\rangle\equiv\langle\hat{\sigma}_{\mu\nu}\rangle$,
 $\langle\hat{\sigma}_{\mu\nu}^{(j)}\hat{a}\rangle=\langle\hat{\sigma}_{\mu\nu}^{(1)}\hat{a}\rangle\equiv\langle\hat{\sigma}_{\mu\nu}\hat{a}\rangle$,
and $\langle\hat{\sigma}_{\mu\nu}^{(i)}\hat{\sigma}_{\mu^{\prime}\nu^{\prime}}^{(j)}\rangle_{i\neq j}=\langle\hat{\sigma}_{\mu\nu}^{(1)}\hat{\sigma}_{\mu^{\prime}\nu^{\prime}}^{(2)}\rangle\equiv\langle\hat{\sigma}_{\mu\nu}\hat{\sigma}_{\mu^{\prime}\nu^{\prime}}\rangle$ ($\mu,\nu={\rm S}, {\rm P}, x, g;\   i,j=1,\dots,N$).

Starting from the average photon number $\langle\hat{a}^{\dagger}\hat{a}\rangle$,
we derive a series of dynamical equations of operator expectation
values until they are closed, which are given in Appendix~\ref{AppA}. Then, we can obtain the steady-state photon number $\langle\hat{a}^{\dagger}\hat{a}\rangle_{\rm s}$ and laser power $P\equiv\kappa\omega_c\langle\hat{a}^{\dagger}\hat{a}\rangle_{\text{s}}$ by numerically solving the above dynamical equations.

For each Raman intensity $\widetilde{\Omega}$ and Raman ratio $\Omega_{\alpha}/\Omega_{\beta}$, the steady-state photon number $\langle\hat{a}^{\dagger}\hat{a}\rangle_{\rm s}$ is a function of the pumping rate $\eta$. As shown in Fig.~\ref{fig2}, in some regions of $\eta$, $\langle\hat{a}^{\dagger}\hat{a}\rangle_{\rm s}$ is much larger than unit (e.g., $10^3$--$10^6$). Outside these regions,
$\langle\hat{a}^{\dagger}\hat{a}\rangle_{\rm s}$ is only of the order of unit or even below. Moreover, at the border of these regions, $\langle\hat{a}^{\dagger}\hat{a}\rangle_{\rm s}$ sharply increases or decreases with $\eta$. Apparently, the laser is generated in the regions with large $\langle\hat{a}^{\dagger}\hat{a}\rangle_{\rm s}$. Hence, we define the value of $\eta$ at the border of these regions as the laser threshold \footnote{Precisely speaking, the border of the regions with large $\langle\hat{a}^{\dagger}\hat{a}\rangle_{\rm s}$ is not a single point in the $\eta$ axis, but has finite width, as shown in Fig.~\ref{fig2}. Here we just choose one point in the border to be the pumping threshold, and our result is not influenced by the chosen of this point.}. 

\subsection{Calculation of Lasing Frequency and Linewidth}

\begin{figure}[t]
\centering{}\includegraphics[scale=0.5]{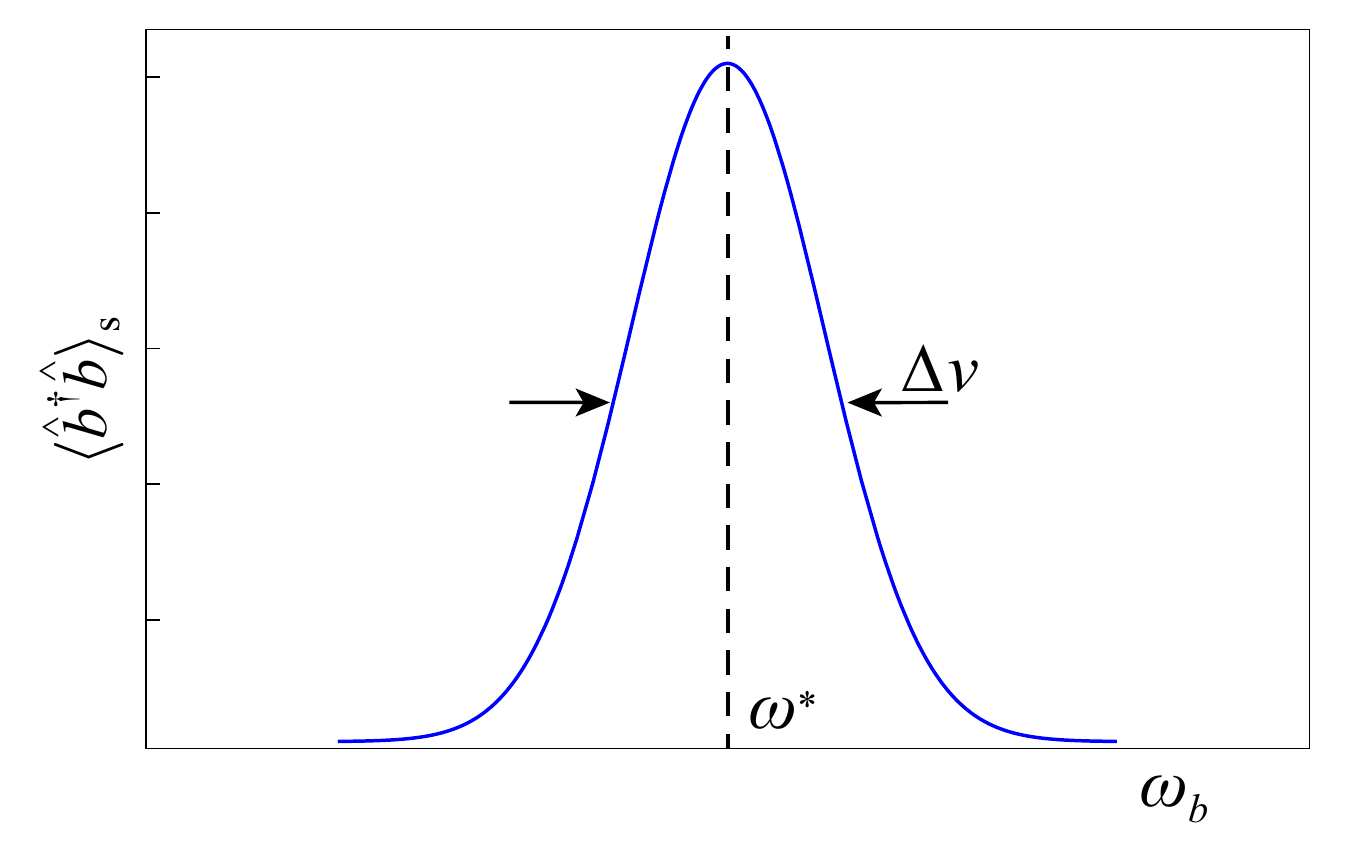}
\caption{\label{fig3}The schematic diagram of the spectrum of filter cavity. The laser frequency and linewidth correspond to the central frequency and FWHM of this spectrum.}
\end{figure}

One can not  obtain the laser spectrum directly from the steady-state solutions of Eq.~(\ref{eq:master}). Here, we employ the filter-cavity method~\citep{Debnath2018,Zhang2021} to calculate the frequency and linewidth of the output laser numerically. Specifically, we assume that a low-dissipation ``filter cavity" is weakly coupled with the laser cavity, which is described by the Hamiltonian
\begin{equation}
 \hat{H}_{f}=\omega_{b}\hat{b}^{\dagger}\hat{b}+\zeta(\hat{b}^{\dagger}\hat{a}+\hat{a}^{\dagger}\hat{b}).   
\end{equation}
Here, $\hat{b}$ is the photon annihilation operator of the filter cavity with angular frequency $\omega_{b}$, and $\zeta$ denotes its coupling strength with the lasing cavity. As both the dissipation rate of the filter cavity and its coupling with the laser mode are assumed to be much smaller than the Rabi frequencies and the dissipation rates of the lasing system, the filter cavity has negligible influence on the lasing process. Meanwhile, the laser photon can enter into and dissipate from the filter cavity. Therefore, the spectrum of the steady-state photon number $\langle\hat{b}^{\dagger}\hat{b}\rangle_{\text{s}}$ versus the filter cavity frequency $\omega_{b}$ will provide information about the laser frequency and linewidth of our scheme. As shown in Fig.~\ref{fig3}, the spectrum of $\langle\hat{b}^{\dagger}\hat{b}\rangle_{\text{s}}$ has a single peak whose central frequency is just the lasing angular frequency $\omega^\ast$. The full width at half maximum (FWHM) of this peak corresponds to the laser linewidth $\Delta\nu$~\citep{Debnath2018}.

Before we present the numerical results, it is worth mentioning that the laser linewidth $\Delta\nu$ can also be obtained by the quantum regression theorem~\citep{Lax1963,opensystem2002,Meiser2009}. Though the quantum-regression method is not as efficient as the filter-cavity one for numerical calculation, it can give an approximate expression of the laser linewidth in terms of $\langle\hat{\sigma}_{\mu\nu}\rangle_{\rm s}\ (\mu,\nu={\textrm S}, {\textrm P}, x, g$) as
\begin{widetext}
\begin{equation}
\Delta\nu =\left|\frac{\kappa+\frac{N\Omega_{c}^{2}}{F}[\eta\Omega_{\alpha}\text{Im}\langle\hat{\sigma}_{\text{S}x}\rangle_{\mathrm{s}}+\Omega_{\alpha}\Omega_{\beta}\text{Re}\langle\hat{\sigma}_{\text{P}x}\rangle_{\mathrm{s}}-(\eta\Gamma+\Omega_{\beta}^{2})\langle\hat{\sigma}_{xx}-\hat{\sigma}_{gg}\rangle_{\mathrm{s}}]}{
1+\frac{\kappa}{F}[\eta\Gamma+(\gamma_{0}+\eta)(\Gamma+\eta)+\Omega_{\alpha}^{2}+\Omega_{\beta}^{2}]+\frac{N\Omega_{c}^{2}}{F}[\Omega_{\alpha}\text{Im}\langle\hat{\sigma}_{\text{S}x}\rangle_{\mathrm{s}}-(\Gamma+\eta)\langle\hat{\sigma}_{xx}-\hat{\sigma}_{gg}\rangle_{\mathrm{s}}]}\right|, \label{alw}
\end{equation}
\end{widetext}
where we have defined the notations $\Gamma=\gamma_{x}+\gamma_{\textrm{P}}+\eta$ and $F={\eta\Omega_{\alpha}^{2}+(\gamma_{0}+\eta)(\eta\Gamma+\Omega_{\beta}^{2})}$.
The detailed derivation of Eq.~(\ref{alw}) is given in Appendix~\ref{AppB}, which also shows that Eq.~(\ref{alw}) coincides with the numerical result very well. We will discuss the effects of the dark-bright state coherence with the help of Eq.~(\ref{alw}) in Sec.~\ref{sec:V}.

\section{\label{lp} Laser Properties}

In this section, we consider an ensemble of cold alkaline-earth-metal $^{88}\mathrm{Sr}$ atoms as the gain medium to 
illustrate the properties of the superradiant laser generated via our scheme. The laser using the same kind of atoms, but without the Raman beams, has been experimentally achieved in the superradiant crossover regime~\citep{Norcia2016}. In our following second-order mean-field calculation, we adopt the parameters achievable in experiments, for instance, $N=10^5$, $\kappa=2\pi\times150\mathrm{kHz}$. The spontaneous decay rates of the $^{88}\mathrm{Sr}$ atom are
$\gamma_{x}=2\pi\times2.6\mathrm{MHz}$, $\gamma_{\textrm{P}}=2\pi\times1.8\mathrm{MHz}$, and $\gamma_{0}=2\pi\times7.5\mathrm{kHz}$~\citep{Courtillot2005}.

\subsection{Power and Linewidth}
We show the superradiant laser power $P$ and the linewidth $\Delta\nu$ as functions of the pumping rate $\eta$ in Fig.~\ref{fig2}. The black-solid (red-dashed) lines represent the case with the Raman strength $\widetilde{\Omega}=2\pi\times 10^{1/2}\textrm{MHz}\ (2\pi\times 10\textrm{MHz})$, and the subfigures from left column to right are plotted for the Raman ratio $\Omega_\alpha/\Omega_\beta=1$, $10^{1/2}$ and $10$. We see that the power $P$ increases with the pumping rate $\eta$ in the lasing regime, which is similar to other superradiant systems. However, in the pumping-linewidth curve, while only one minimum appears in the system without Raman beams~\citep{Meiser2009,Debnath2018}, another local minimum point emerges in our scheme to the left of the former. Moreover, for proper Raman strengths and ratios, the new local minimum of linewidth becomes the global minimum with the pumping rate smaller than  $2\pi\times10$kHz. This may inhibit the heating effect in the lasing process, which is helpful for the continuous output. Hence, the emergence of this new local minimum implies that our scheme may generate a narrow-linewidth ($2\pi\times 1$Hz) laser with a considerable power ($10^{-10}$W)  for practical use. For clarity, we mark the regions satisfying the conditions
\begin{eqnarray}
P &\geq& 10^{-10}{\textrm W},\label{c1}\\
\Delta\nu &\leq& 2\pi\times 1{\textrm{Hz}},\label{c2}\\
\eta &\leq& 2\pi\times10{\textrm{kHz}}.\label{c3}
\end{eqnarray}
in purple in Fig.~\ref{fig2}.

We emphasize that the double-minimum behavior of linewidth is a characteristic resulting from the  Raman-induced coherence, which is a significant difference of our superradiant lasing scheme from the previous works without Raman beams~\citep{Meiser2009,Debnath2018,Holland2017}. We leave the discussion of the underlying physics in Sec.~\ref{sec:V}.

\begin{figure}[t]
\centering{}\includegraphics[scale=0.34]{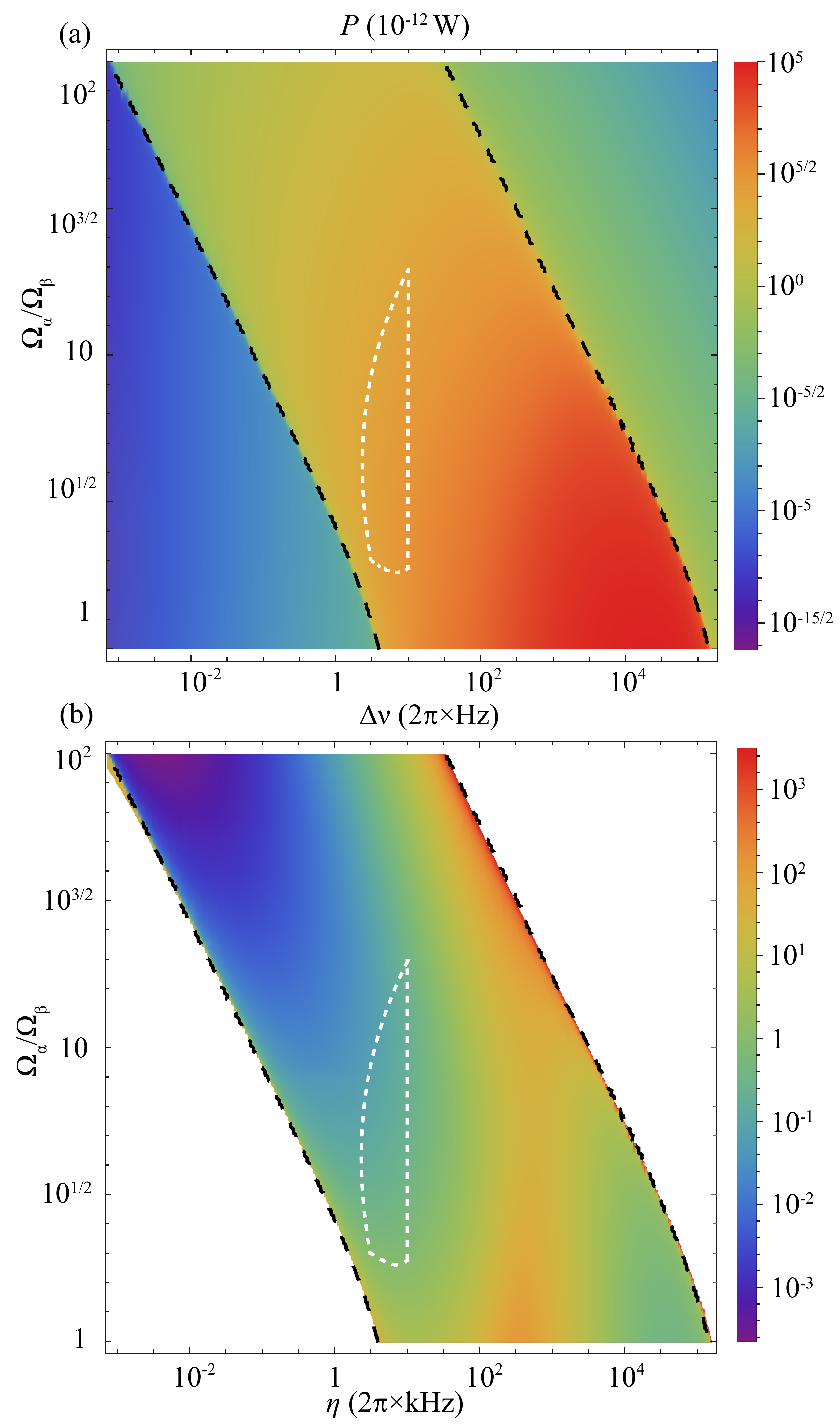}
\caption{\label{fig4}2D plots of {\bf (a)} the power $P$ and {\bf (b)} the linewidth $\Delta\nu$ as functions of the pumping rate $\eta$ and the Raman ratio $\Omega_{\alpha}/\Omega_{\beta}$. The thresholds of the laser scheme are illustrated by black-dashed lines, and the laser is generated in the regions in between. The area surrounded by the white-dashed line represents the region satisfying $P\protect\geq 10^{-10}$W, $\Delta\nu\protect\leq2\pi\times1$Hz, and $\eta\protect\leq2\pi\times10$kHz. Here, we set $\widetilde{\Omega}=2\pi\times10^{1/2}$MHz. In {\bf (b)}, we only show the linewidth in the lasing region.}
\end{figure}

In Fig.~\ref{fig4}, we provide more comprehensive information about our lasing scheme by two-dimensional plots of the laser power $P$ and linewidth $\Delta\nu$ as functions of the pumping rate $\eta$ and the Raman ratio $\Omega_{\alpha}/\Omega_{\beta}$. The Raman strength $\widetilde{\Omega}$ is fixed at $2\pi\times10^{1/2}$MHz. The black-dashed lines represent the threshold of our lasing scheme, and the area surrounded by the white-dashed line shows the region satisfying Eqs.~(\ref{c1}-\ref{c3}). In Fig.~\ref{fig4}(b), we only plot the laser linewidth inside the lasing region. 

Figure~\ref{fig4} shows that the Raman ratio $\Omega_\alpha/\Omega_\beta$ can be used to tune the laser power and linewidth. When increasing the Raman ratio, we find that both the laser power and linewidth, working around the left local minimum of linewidth, decrease continuously. Thus, there is competition between achieving considerable output power and keeping small linewidth. In particular, when $\Omega_\alpha/\Omega_\beta$ is raised to the order of $10^2$, we can realize the superradiant laser with millihertz linewidth and  $10^{-12}$W output power. This linewidth is at least six orders of magnitude smaller than the natural linewidth of the $^3$P$_1$ state of $^{88}$Sr and comparable with that of the $^3$P$_0$ state of $^{87}$Sr~\citep{Meiser2009}.

\begin{figure}[t]
\begin{centering}
\includegraphics[scale=0.18]{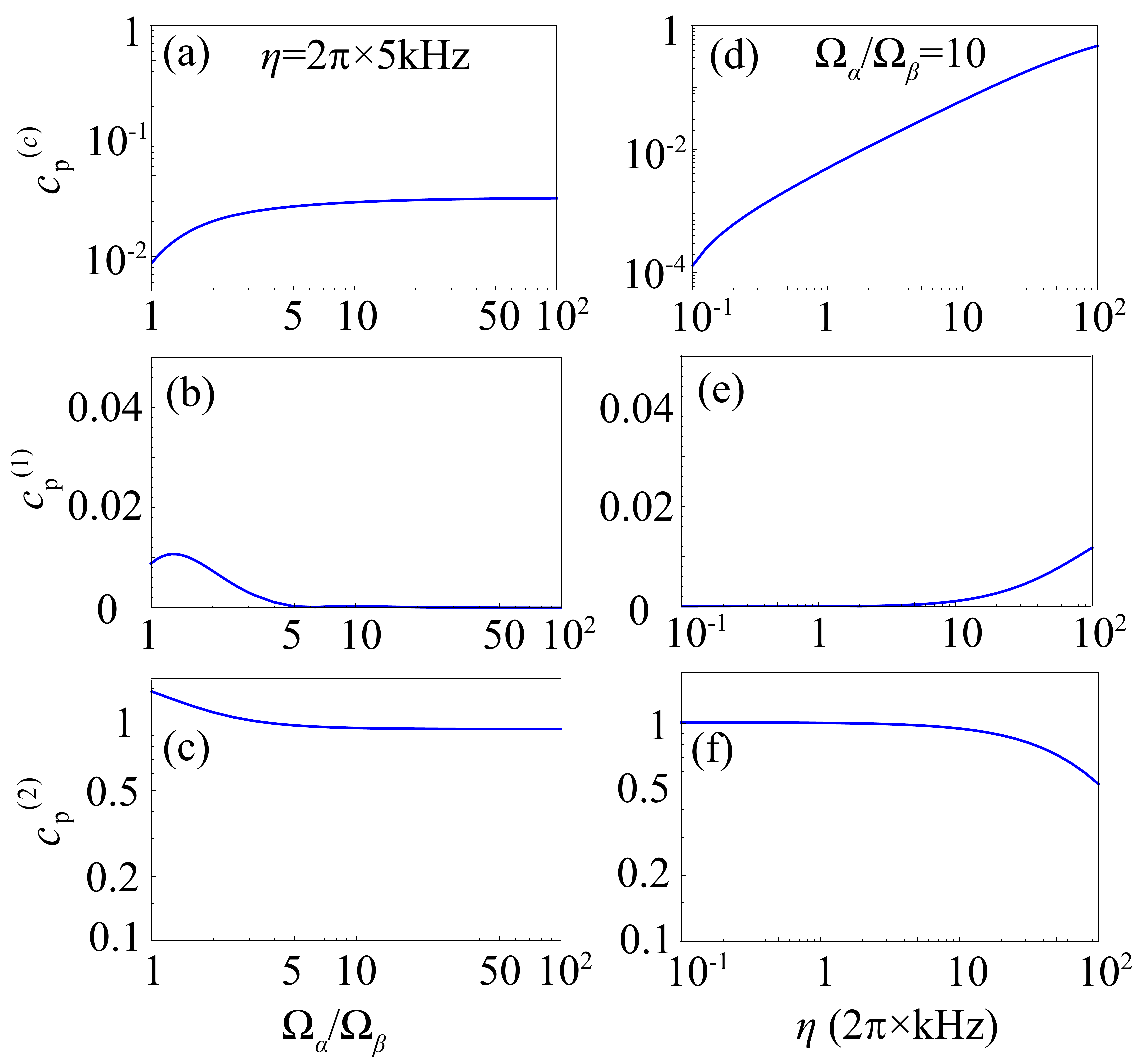}
\par\end{centering}
\caption{\label{fig5}The pulling coefficients $c_{\text{p}}^{(c)}$, $c_{\text{p}}^{(1)}$,
and $c_{\text{p}}^{(2)}$ as functions of the Raman ratio $\Omega_\alpha/\Omega_\beta$ {\bf (a-c)}
with the pumping rate $\eta=2\pi\times5$kHz, and as functions of
the pumping rate $\eta$ {\bf (d-f)} with the Raman ratio $\Omega_\alpha/\Omega_\beta=10$.
Here, we set $\widetilde{\Omega}=2\pi\times10^{1/2}$MHz.}
\end{figure}

\subsection{Lasing Frequency and Pulling Coefficients}

When the cavity mode is resonant with the atomic $^1$S$_0$-to-$^3$P$_1$ transition ($\delta_c=0$), and both of the two Raman beams are resonant with the corresponding
atomic transitions ($\delta_\alpha=\delta_\beta=0$), numerical calculation shows that the central frequency $\omega^{\ast}$ of the output laser equals the atomic $^1$S$_0$-to-$^3$P$_1$ transition
frequency $\omega_0$, i.e., $\omega^{\ast}=\omega_0$.

Nevertheless, in realistic systems, fluctuations of the frequencies of the cavity mode and Raman beams (non-zero $\delta_{c,\alpha,\beta}$) will shift the lasing frequency from $\omega_0$. The following three pulling coefficients can describe the stability of the lasing frequency under these fluctuations,
\begin{eqnarray}
c_{\text{p}}^{(i)}  =\left|\frac{\partial\delta^{\ast}}{\partial\delta_{i}}\right|_{\delta_c=\delta_1=\delta_2=0}\ \ \ \ (i=c,1,2),
\end{eqnarray}
where $\delta^{\ast}\equiv\omega^{\ast}-\omega_{0}$ is the fluctuation of the laser frequency, and  $\delta_{1}\equiv\delta_{\alpha}+\delta_{\beta}$ ($\delta_{2}\equiv\delta_{\alpha}-\delta_{\beta}$) is the one-photon (two-photon) detuning. According to the definition of $\delta_{\alpha}$ and $\delta_{\beta}$ in Eqs.~(\ref{da}) and (\ref{db}), the detunings $\delta_{1}$ and $\delta_{2}$ are determined by the summation and difference of the frequencies of the two Raman beams (i.e., $\omega_\alpha+\omega_\beta$ and $\omega_\alpha-\omega_\beta$), respectively.

In Fig.~\ref{fig5}, we plot the pulling coefficients $c_{\mathrm{p}}^{(c)}$, $c_{\mathrm{p}}^{(1)}$, and $c_{\mathrm{p}}^{(2)}$ as functions of
the Raman ratio $\Omega_{\alpha}/\Omega_{\beta}$ and pumping rate $\eta$. Our results show that for $\eta\sim 2\pi\times5$kHz, we have $c_p^{(c)}\sim 10^{-2}$, $c_p^{(1)}\approx 0$, and $c_p^{(2)}\approx 1$. Therefore, in this region,
the central frequency of the output laser is robust against the fluctuation of the cavity frequency and that of the frequency sum of the two Raman beams. However, the fluctuation of the frequency difference of the two Raman beams $\delta_{2}$ results in an uncertainty of the central lasing frequency which approximately equals to $\delta_2$. In the current experiments, via locking the two Raman beams with an optical comb or two modes of the same cavity~\citep{Fortier2019},
one can suppress $|\delta_{2}|$ to the level of hertz (or even lower). Therefore, the uncertainty of the central lasing frequency corresponding to $\delta_2$ can approach the order of hertz as well.

\section{\label{sec:V}Coherence induced double minima of linewidth}

In the above sections, we have demonstrated that the appearance of double minima in the pumping-linewidth curve is a crucial characteristic of our lasing scheme. Based on this fact, it is possible to realize a laser with a relatively small linewidth and large power at a low pumping rate (e.g., $\Delta\nu\leq 2\pi \times 1$Hz, $P \geq 10^{-10}$W at $\eta \leq 2\pi \times 10$kHz). In this section, we reveal that such double-minimum feature stems from the Raman-induced coherence between the dark state $|{\rm D}\rangle$ and the bright state $|{\rm B}\rangle$. To this end, we illustrate the effect of this coherence from two aspects. First, we show that a simplified three-level model (TLM) cannot fully capture the features of the four-level lasing system; then, a rescaled coherence measure is defined for quantitative investigation.

\subsection{\label{sec:Va}Three-Level Models}

In our system, as shown in Fig.~\ref{fig1}(d), the atoms pumped to the state $|{\rm S}\rangle$ can transit to both of the
states $|{\rm D}\rangle$ and $|{\rm B}\rangle$, and then emit laser photons to the cavity mode. However, since the state
$|{\rm S}\rangle$ is coupled to $|{\rm B}\rangle$ by the Raman beams, the atoms in state $|{\rm B}\rangle$ can be ``re-pumped" to $|{\rm S}\rangle$. The atoms in state $|{\rm D}\rangle$, in contrast, cannot directly transit back to $|{\rm S}\rangle$ and have a one-hundred percent possibility to emit laser photons. The above facts indicate that it is the dark state rather than the bright one that dominates in the lasing process. Thus, a simplified TLM ignoring the state $|{\rm B}\rangle$ may capture certain main properties, such as power and threshold, of our laser scheme. 

\begin{figure}[t]
\begin{centering}
\includegraphics[scale=0.6]{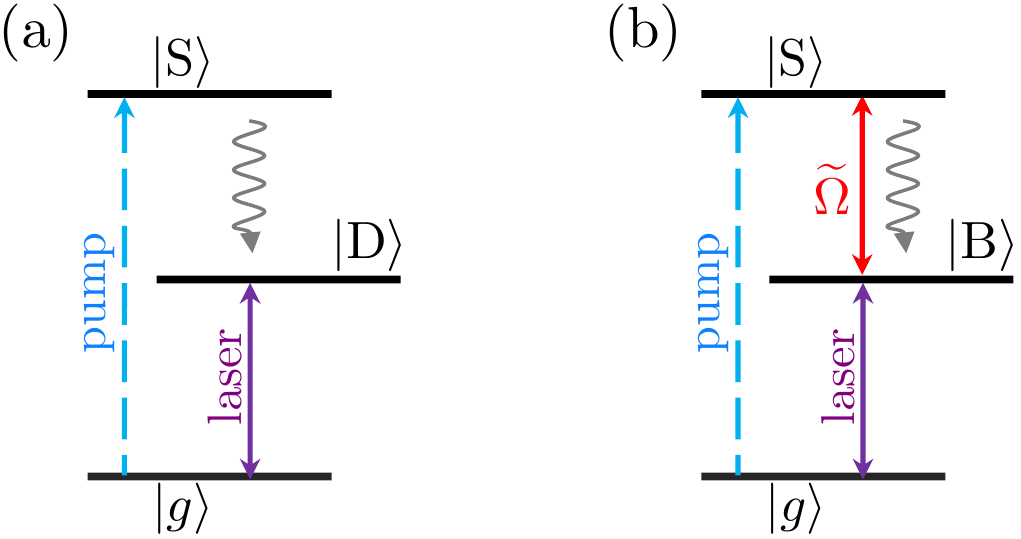}
\par\end{centering}
\caption{\label{fig6}The simplified three-level models used in the Sec.~\ref{sec:Va}. {\bf (a)}: The dark three-level model contains the ground state $|g\rangle$, the $^1\mathrm{S}_0$ state $|{\rm S}\rangle$, and the dark state $|{\rm D}\rangle$. {\bf (b)}: The bright three-level model contains the bright state $|{\rm B}\rangle$ instead of $|{\rm D}\rangle$. In addition, an effective transition induced by the Raman beams couples the states $|{\rm B}\rangle$ with $|{\rm S}\rangle$.}
\end{figure}

In order to verify this thinking, we investigate a simplified TLM containing the atomic states $|g\rangle$, $|{\rm D}\rangle$ and $|{\rm S}\rangle$, which is named as ``dark TLM" and schematically shown in Fig.~\ref{fig6}(a). Compared to the four-level model shown in Fig.~\ref{fig1}(d), the dark TLM ignores the bright state $|{\rm B}\rangle$ and the spontaneous-decay-induced coupling between $|{\rm D}\rangle$ and $|{\rm B}\rangle$. As the dark state $|{\rm D}\rangle$ is a superposition of the atomic states $|x\rangle$ and $|{\rm P}\rangle$ (see Eq.~(\ref{dark})), the parameters used here is also superposed by that of the states $|x\rangle$ and $|{\rm P}\rangle$. For instance, the spontaneous decay rates of state $|{\rm S}\rangle$ to $|{\rm D}\rangle$ and $|{\rm D}\rangle$ to $|g\rangle$ are ($\Omega^{2}_{\alpha}\gamma_{{\rm P}}+\Omega^{2}_{\beta}\gamma_{x}$)/$\widetilde{\Omega}^{2}$ and $\Omega^{2}_{\beta}\gamma_{0}$/$\widetilde{\Omega}^{2}$. The Rabi frequency of the coupling between the state  $|{\rm D}\rangle$ and the cavity mode is $\Omega_{\beta}\Omega_{c}$/$\widetilde{\Omega}$.

\begin{figure}[t]
\centering{}\includegraphics[scale=0.35]{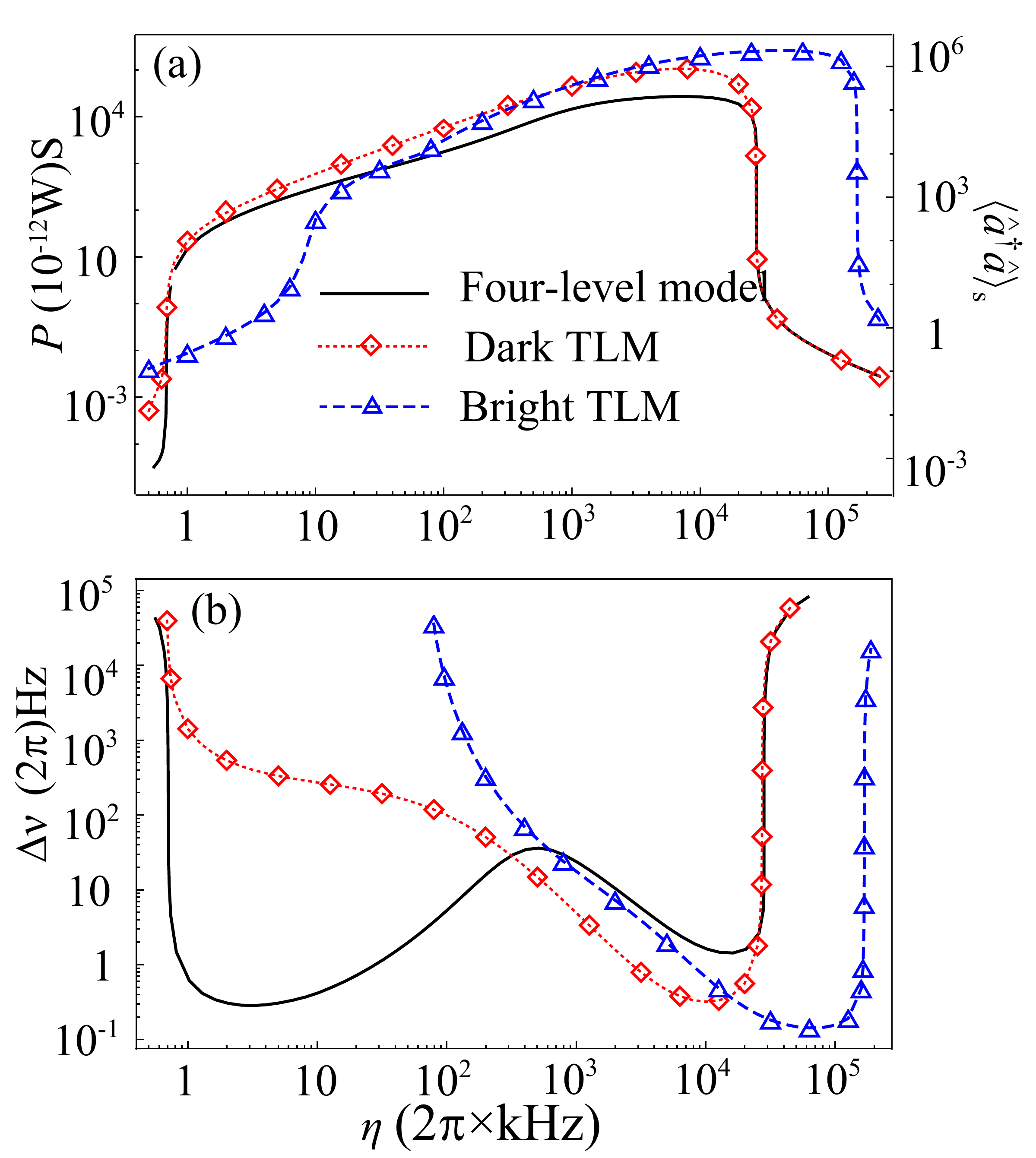}
\caption{\label{fig7}Comparison of {\bf (a)} the laser output power and {\bf (b)} linewidth of the four-level lasing model in Sec.~\ref{sec:model} (black lines), the dark TLM (red-dotted lines with diamonds), and the bright TLM (blue-dashed lines with triangles). Here, we set $\Omega_{\alpha}/\Omega_{\beta}=10^{1/2}$ and $\widetilde{\Omega}=2\pi\times10^{1/2}$MHz.}
\end{figure}

We calculate the output power of the dark TLM with the second-order mean-field method and compare the result with that of the four-level scheme given in Sec.~\ref{lp}. As shown in Fig.~\ref{fig7}, the laser power and threshold of the dark TLM are very close to that of the four-level model. Nevertheless, the laser linewidth of the dark TLM has only one minimum at about $\eta=2\pi\times 10^4$kHz, while another local minimum emerges in the four-level model at around $\eta=2\pi\times 3$kHz. Moreover, near the new local minimum, the linewidth of the four-level model is of the subhertz level, and three orders of magnitude smaller than that of the dark TLM. The above results yield that the dark lasing state alone cannot explain all the properties of the four-level model, especially the double-minimum behavior of the linewidth as a function of the pumping rate.

As a comparison, we also investigate a ``bright TLM" which contains the atomic states $|g\rangle$, $|{\rm B}\rangle$ and $|{\rm S}\rangle$, as shown in Fig.~\ref{fig6}(b). Unlike the dark TLM, in the bright TLM, the $|{\rm B}\rangle$ state coherently couples to the $|{\rm S}\rangle$ state with an effective coupling strength $\widetilde{\Omega}$ induced by the Raman beams. The other parameters are similar to those in the dark TLM. In Fig.~\ref{fig7}, the blue-dashed lines with triangles show the power and linewidth of the bright TLM, which are much different from those of the four-level model. 

Clearly, neither the dark state nor the bright one alone is the reason for the double-minimum linewidth of the four-level model. The failure of these two TLMs suggests that the coherence between $|{\rm D}\rangle$ and $|{\rm B}\rangle$ may play a significant role in the double-minimum behavior.

\subsection{Coherence between the dark and bright states}

\begin{figure}[t]
\centering{}\includegraphics[scale=0.4]{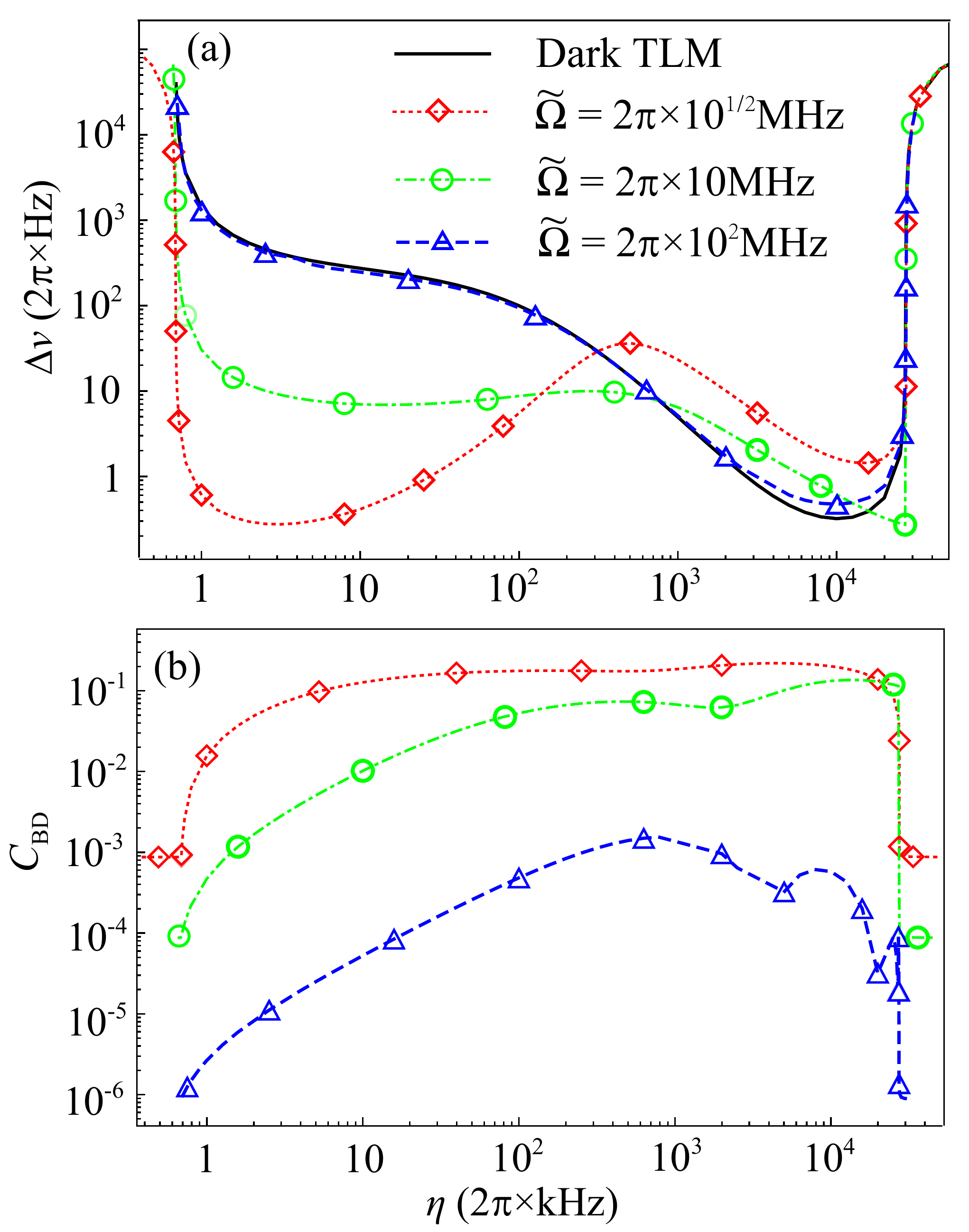}
\caption{\label{fig8}{\bf (a)} The linewidth $\Delta\nu$ and {\bf (b)} the rescaled coherence $C_{{\rm B}{\rm D}}$ of the four-level lasing model introduced in Sec.~\ref{sec:model}. The results are plotted with $\widetilde{\Omega}=2\pi\times10^{1/2}$MHz (red-dotted line with diamonds), $\widetilde{\Omega}=2\pi\times10$MHz (green-dotdashed line with circles), and $2\pi\times10^{2}$MHz (blue-dashed line with triangles). The linewidth of the dark TLM is also displayed with the black-solid line for comparison. Here, we have chosen the Raman ratio $\Omega_{\alpha}/\Omega_{\beta}=10^{1/2}$.}
\end{figure}

The steady-state expectation value $\langle\hat{\sigma}_{{\rm B}{\rm D}}\rangle_{\textrm{s}}\equiv \textrm{Tr}[\vert \textrm{B}\rangle\langle\textrm{D}\vert\hat{\rho}(\infty)]$ naturally measures the coherence between the dark and bright states. When rewriting the linewidth equation Eq.~(\ref{alw}) with the basis of the states $\vert \textrm{D}\rangle$ and $\vert \textrm{B}\rangle$, we can find that the term $\langle\hat{\sigma}_{{\rm B}{\rm D}}\rangle_{\textrm{s}}$ explicitly appears in the expression of $\Delta\nu$.
However, the direct comparison of $\langle\hat{\sigma}_{{\rm B}{\rm D}}\rangle_{\textrm{s}}$ between different parameter cases is meaningless because the populations in the dark and bright states vary with $\eta$ and $\Omega_{\alpha(\beta)}$. Therefore, we define a rescaled measure of coherence as
\begin{eqnarray}
C_{{\rm B}{\rm D}}=\frac{|\langle\hat{\sigma}_{{\rm B}{\rm D}}\rangle_{\textrm{s}}|}{\langle\hat{\sigma}_{{\rm D}{\rm D}}\rangle_{\textrm{s}}
+\langle\hat{\sigma}_{{\rm B}{\rm B}}\rangle_{\textrm{s}}},\label{eq:CBD}
\end{eqnarray}
where $\langle\hat{\sigma}_{{\rm D}{\rm D}}\rangle_{\textrm{s}}$ and $\langle\hat{\sigma}_{{\rm B}{\rm B}}\rangle_{\textrm{s}}$ are the steady-state populations of the dark and bright states, respectively.

In Fig.~\ref{fig8}, we plot the laser linewidth and the corresponding coherence measure $C_{{\rm B}{\rm D}}$ with different values of $\widetilde{\Omega}$ as a function of the pumping rate $\eta$. Here, we choose the same value of $\Omega_{\alpha}/\Omega_{\beta}=10^{1/2}$ as in Fig.~\ref{fig7}.
It can be seen from Fig.~\ref{fig8}(a) that the double-minimum behavior of $\Delta\nu$ is more obvious for small Raman strength  $\widetilde{\Omega}$. When we increase the Raman strength $\widetilde{\Omega}$ from $2\pi\times 10^{1/2}$MHz to $2\pi\times 10$MHz, the left local minimum of linewidth increases from below $2\pi\times 1$Hz to $2\pi\times 10$Hz. Further increasing the Raman strength to $2\pi\times 10^{2}$MHz, we see that the double-minimum curve reduces to a single-minimum one which coincides with that of the dark TLM. Meanwhile, the rescaled coherence $C_{{\rm B}{\rm D}}$ corresponding to the left local minimum of linewidth decreases from the order of $10^{-1}$ to $10^{-5}$ as $\widetilde{\Omega}$ increases from $2\pi\times 10^{1/2}$MHz to $2\pi\times 10^{2}$MHz (see Fig.~\ref{fig8}(b)). Such a correlation between a large $C_{{\rm B}{\rm D}}$ and the  double-minimum behavior of linewidth confirms the importance of the coherence between the dark and bright states in our lasing scheme.\\

\section{Discussions and conclusions\label{sec:conclusions}}

In this work, we have developed a alternative scheme of superradiant laser with Raman transitions. By dressing the $^{3}\textrm{P}_1$ state with the long-lived $^{3}\textrm{P}_0$ state in bosonic alkaline-earth(-like) atoms via two Raman beams, our scheme can reduce the laser linewidth to the hertz level, much smaller than the natural linewidth of the $^{3}\textrm{P}_1$ state. What is more, due to the double-minimum feature of linewidth, our scheme can achieve a laser with narrow linewidth $\lesssim2\pi\times1$Hz and considerable output power $\gtrsim10^{-10}$W ($\sim 10^{3}$ photons in the steady state) at a low pumping rate ($\lesssim2\pi\times 10$kHz). 

In our system, the Raman beams play two critical roles. First,  they are the origin of the steady-state coherence between the dark and bright states which is crucial to the double-minimum behavior of linewidth. Second, they mix the $^{3}\textrm{P}_0$ state into the lasing state with the superposition coefficients tunable. This reduces the laser linewidth significantly. Two simplified three-level lasing models are investigated to reveal the sole effect of the dark and bright states. A rescaled quantity is also introduced to measure the coherence between these two states.   

Moreover, the lasing frequency is robust against the fluctuations of the cavity length and the one-photon detuning. However, a nonzero two-photon detuning will fluctuates the laser frequency with almost the same amount. Fortunately, this two-photon detuning can be well controlled by locking the two Raman beams to an optical comb or two modes of the same cavity.

Our results are obtained with the filter-cavity method and the quantum regression theorem. These two approaches are based on the second-order mean-field master equation and coincide with each other very well. Our work presents a feasible method for realizing the narrow linewidth superradiant laser toward continuous lasing. The properties of this laser, such as threshold, output power, and linewidth, can be tuned by the Rabi frequencies of two Raman beams. Our work greatly improves the output performance of the superradiant laser system with coherence induced by Raman transitions and may offer a firm foundation for its practical use in future.  

\begin{acknowledgments}
GD thank Dr. Jin-Fu Chen for very helpful discussions. GD is supported by NSFC Grant No. 12205211. YY  is supported by NSFC Grant No. 12175204. PZ is supported by National Key Research and Development Program of China Grant No. 2018YFA0306502, and NSAF Grant No. U1930201.   DX is supported by NSFC Grant No. 12075025.
\end{acknowledgments}

\appendix

\section{\label{AppA}The Mean-Field Dynamical Equations}

Using the second-order mean-field theory described in the main text,
a set of closed equations of motion for the expectation values of operators are obtained and listed as follows:
\begin{widetext}
\begin{eqnarray*}
\frac{d}{dt}\langle\hat{a}^{\dagger}\hat{a}\rangle &=&-\kappa\langle\hat{a}^{\dagger}\hat{a}\rangle+i\frac{N\Omega_{c}}{2}\left(\langle\hat{\sigma}_{xg}\hat{a}\rangle-\text{h.c.}\right),\\
\frac{d}{dt}\langle\hat{\sigma}_{xg}\hat{a}\rangle 
&=&-[i\delta_{c}+\frac{1}{2}(\gamma_{0}+\eta+\kappa)]\langle\hat{\sigma}_{xg}\hat{a}\rangle+i\frac{\Omega_{\alpha}}{2}\langle\hat{\sigma}_{\text{S}g}\hat{a}\rangle-i\frac{\Omega_{c}}{2}[\langle\hat{\sigma}_{xx}\rangle+\langle\hat{a}^{\dagger}\hat{a}\rangle\left(\langle\hat{\sigma}_{xx}\rangle-\langle\hat{\sigma}_{gg}\rangle\right)+\left(N-1\right)\langle\hat{\sigma}_{xg}\hat{\sigma}_{gx}\rangle],\\
\frac{d}{dt}\langle\hat{\sigma}_{\text{P}g}\hat{a}\rangle &=&[i(\delta_{\beta}-\delta_{c})-\frac{1}{2}(\eta+\kappa)]\langle\hat{\sigma}_{\text{P}g}\hat{a}\rangle+i\frac{\Omega_{\beta}}{2}\langle\hat{\sigma}_{\text{S}g}\hat{a}\rangle-i\frac{\Omega_{c}}{2}[\langle\hat{a}\hat{a}^{\dagger}\rangle\langle\hat{\sigma}_{\text{P}x}\rangle+\left(N-1\right)\langle\hat{\sigma}_{\text{P}g}\hat{\sigma}_{gx}\rangle],\\
\frac{d}{dt}\langle\hat{\sigma}_{\text{S}g}\hat{a}\rangle &=&[i(\delta_{\alpha}-\delta_{c})-\frac{1}{2}(\Gamma+\kappa)]\langle\hat{\sigma}_{\text{S}g}\hat{a}\rangle+i\frac{\Omega_{\alpha}}{2}\langle\hat{\sigma}_{xg}\hat{a}\rangle+i\frac{\Omega_{\beta}}{2}\langle\hat{\sigma}_{\text{P}g}\hat{a}\rangle-i\frac{\Omega_{c}}{2}[\langle\hat{a}\hat{a}^{\dagger}\rangle\langle\hat{\sigma}_{\text{S}x}\rangle+\left(N-1\right)\langle\hat{\sigma}_{\text{S}g}\hat{\sigma}_{gx}\rangle],\\
\frac{d}{dt}\langle\hat{\sigma}_{xg}\hat{\sigma}_{gx}\rangle &=&-(\gamma_{0}+\eta)\langle\hat{\sigma}_{xg}\hat{\sigma}_{gx}\rangle+(i\frac{\Omega_{\alpha}}{2}\langle\hat{\sigma}_{\text{S}g}\hat{\sigma}_{gx}\rangle+\text{h.c.})+(i\frac{\Omega_{c}}{2}\langle\hat{\sigma}_{xg}\hat{a}\rangle+\text{h.c.})(\langle\hat{\sigma}_{xx}\rangle-\langle\hat{\sigma}_{gg}\rangle),\\
\frac{d}{dt}\langle\hat{\sigma}_{xg}\hat{\sigma}_{g\text{P}}\rangle &=&[-i\delta_{\beta}-\frac{1}{2}(\gamma_{0}+2\eta)]\langle\hat{\sigma}_{xg}\hat{\sigma}_{g\text{P}}\rangle+i\frac{\Omega_{\alpha}}{2}\langle\hat{\sigma}_{\text{S}g}\hat{\sigma}_{g\text{P}}\rangle-i\frac{\Omega_{\beta}}{2}\langle\hat{\sigma}_{xg}\hat{\sigma}_{g\text{S}}\rangle +i\frac{\Omega_{c}}{2}[\langle\hat{\sigma}_{xg}\hat{a}\rangle\langle\hat{\sigma}_{x\text{P}}\rangle-\langle\hat{a}^{\dagger}\hat{\sigma}_{g\text{P}}\rangle(\langle\hat{\sigma}_{xx}\rangle-\langle\hat{\sigma}_{gg}\rangle)],\\
\frac{d}{dt}\langle\hat{\sigma}_{xg}\hat{\sigma}_{g\text{S}}\rangle &=&[-i\delta_{\alpha}-\frac{1}{2}(\gamma_{0}+\Gamma+\eta)]\langle\hat{\sigma}_{xg}\hat{\sigma}_{g\text{S}}\rangle+i\frac{\Omega_{\alpha}}{2}(\langle\hat{\sigma}_{\text{S}g}\hat{\sigma}_{g\text{S}}\rangle-\langle\hat{\sigma}_{xg}\hat{\sigma}_{gx}\rangle)-i\frac{\Omega_{\beta}}{2}\langle\hat{\sigma}_{xg}\hat{\sigma}_{g\text{P}}\rangle\\
 && -i\frac{\Omega_{c}}{2}\langle\hat{a}^{\dagger}\hat{\sigma}_{g\text{S}}\rangle(\langle\hat{\sigma}_{xx}\rangle-\langle\hat{\sigma}_{gg}\rangle)+i\frac{\Omega_{c}}{2}\langle\hat{\sigma}_{x\text{S}}\rangle\langle\hat{\sigma}_{xg}\hat{a}\rangle,\\
\frac{d}{dt}\langle\hat{\sigma}_{\text{P}g}\hat{\sigma}_{g\text{P}}\rangle &=&-\eta\langle\hat{\sigma}_{\text{P}g}\hat{\sigma}_{g\text{P}}\rangle+(i\frac{\Omega_{c}}{2}\langle\hat{\sigma}_{\text{P}g}\hat{a}\rangle\langle\hat{\sigma}_{x\text{P}}\rangle-i\frac{\Omega_{\beta}}{2}\langle\hat{\sigma}_{\text{P}g}\hat{\sigma}_{g\text{S}}\rangle+\text{h.c.}),\\
\frac{d}{dt}\langle\hat{\sigma}_{\text{P}g}\hat{\sigma}_{g\text{S}}\rangle &=&[i(\delta_{\beta}-\delta_{\alpha})-\frac{1}{2}(\Gamma+\eta)]\langle\hat{\sigma}_{\text{P}g}\hat{\sigma}_{g\text{S}}\rangle-i\frac{\Omega_{\alpha}}{2}\langle\hat{\sigma}_{\text{P}g}\hat{\sigma}_{gx}\rangle+i\frac{\Omega_{\beta}}{2}(\langle\hat{\sigma}_{\text{S}g}\hat{\sigma}_{g\text{S}}\rangle-\langle\hat{\sigma}_{\text{P}g}\hat{\sigma}_{g\text{P}}\rangle)+i\frac{\Omega_{c}}{2}(\langle\hat{\sigma}_{\text{P}g}\hat{a}\rangle\langle\hat{\sigma}_{x\text{S}}\rangle-\langle\hat{\sigma}_{\text{P}x}\rangle\langle\hat{a}^{\dagger}\hat{\sigma}_{g\text{S}}\rangle),\\
\frac{d}{dt}\langle\hat{\sigma}_{\text{S}g}\hat{\sigma}_{g\text{S}}\rangle &=&-\Gamma\langle\hat{\sigma}_{\text{S}g}\hat{\sigma}_{g\text{S}}\rangle+(i\frac{\Omega_{\alpha}}{2}\langle\hat{\sigma}_{xg}\hat{\sigma}_{g\text{S}}\rangle+i\frac{\Omega_{\beta}}{2}\langle\hat{\sigma}_{\text{P}g}\hat{\sigma}_{g\text{S}}\rangle+i\frac{\Omega_{c}}{2}\langle\hat{\sigma}_{\text{S}g}\hat{a}\rangle\langle\hat{\sigma}_{x\text{S}}\rangle+\text{h.c.}),\\
\frac{d}{dt}\langle\hat{\sigma}_{xx}\rangle &=&-\gamma_{0}\langle\hat{\sigma}_{xx}\rangle+\gamma_{x}\langle\hat{\sigma}_{\text{SS}}\rangle+(-i\frac{\Omega_{\alpha}}{2}\langle\hat{\sigma}_{x\text{S}}\rangle+i\frac{\Omega_{c}}{2}\langle\hat{a}^{\dagger}\hat{\sigma}_{gx}\rangle+\text{h.c.}),\\
\frac{d}{dt}\langle\hat{\sigma}_{\text{PP}}\rangle &=&\gamma_{\text{P}}\langle\hat{\sigma}_{\text{SS}}\rangle-(i\frac{\Omega_{\beta}}{2}\langle\hat{\sigma}_{\text{PS}}\rangle+\text{h.c.}),\\
\frac{d}{dt}\langle\hat{\sigma}_{\text{SS}}\rangle &=&-(\gamma_{x}+\gamma_{\mathrm{P}})\langle\hat{\sigma}_{\text{SS}}\rangle+\eta\langle\hat{\sigma}_{gg}\rangle+(i\frac{\Omega_{\alpha}}{2}\langle\hat{\sigma}_{x\text{S}}\rangle+i\frac{\Omega_{\beta}}{2}\langle\hat{\sigma}_{\text{PS}}\rangle+\text{h.c.}),\\
\frac{d}{dt}\langle\hat{\sigma}_{x\text{P}}\rangle &=&(-i\delta_{\beta}-\frac{1}{2}\gamma_{0})\langle\hat{\sigma}_{x\text{P}}\rangle+i\frac{\Omega_{\alpha}}{2}\langle\hat{\sigma}_{\text{SP}}\rangle-i\frac{\Omega_{\beta}}{2}\langle\hat{\sigma}_{x\text{S}}\rangle+i\frac{\Omega_{c}}{2}\langle\hat{a}^{\dagger}\hat{\sigma}_{g\text{P}}\rangle,\\
\frac{d}{dt}\langle\hat{\sigma}_{x\text{S}}\rangle &=&[-i\delta_{\alpha}-\frac{1}{2}(\gamma_{0}+\gamma_{x}+\gamma_{\mathrm{P}})]\langle\hat{\sigma}_{x\text{S}}\rangle+i\frac{\Omega_{\alpha}}{2}(\langle\hat{\sigma}_{\text{SS}}\rangle-\langle\hat{\sigma}_{xx}\rangle)-i\frac{\Omega_{\beta}}{2}\langle\hat{\sigma}_{x\text{P}}\rangle+i\frac{\Omega_{c}}{2}\langle\hat{a}^{\dagger}\hat{\sigma}_{g\text{S}}\rangle,\\
\frac{d}{dt}\langle\hat{\sigma}_{\text{PS}}\rangle &=&[i(\delta_{\beta}-\delta_{\alpha})-\frac{1}{2}(\gamma_{x}+\gamma_{\mathrm{P}})]\langle\hat{\sigma}_{\text{PS}}\rangle-i\frac{\Omega_{\alpha}}{2}\langle\hat{\sigma}_{\text{P}x}\rangle+i\frac{\Omega_{\beta}}{2}(\langle\hat{\sigma}_{\text{SS}}\rangle-\langle\hat{\sigma}_{\text{PP}}\rangle).
\end{eqnarray*}
\end{widetext}
Here, we have used relation $\langle\hat{\sigma}_{gg}\rangle+\langle\hat{\sigma}_{xx}\rangle+\langle\hat{\sigma}_{\text{SS}}\rangle+\langle\hat{\sigma}_{\text{PP}}\rangle=1$.

\section{\label{AppB} Linewidth Obtained by Quantum Regression Theorem}
In this appendix, we show how to obtain the analytical expression of the linewidth via the quantum regression theorem.

\begin{figure}[t]
\begin{centering}
\includegraphics[scale=0.19]{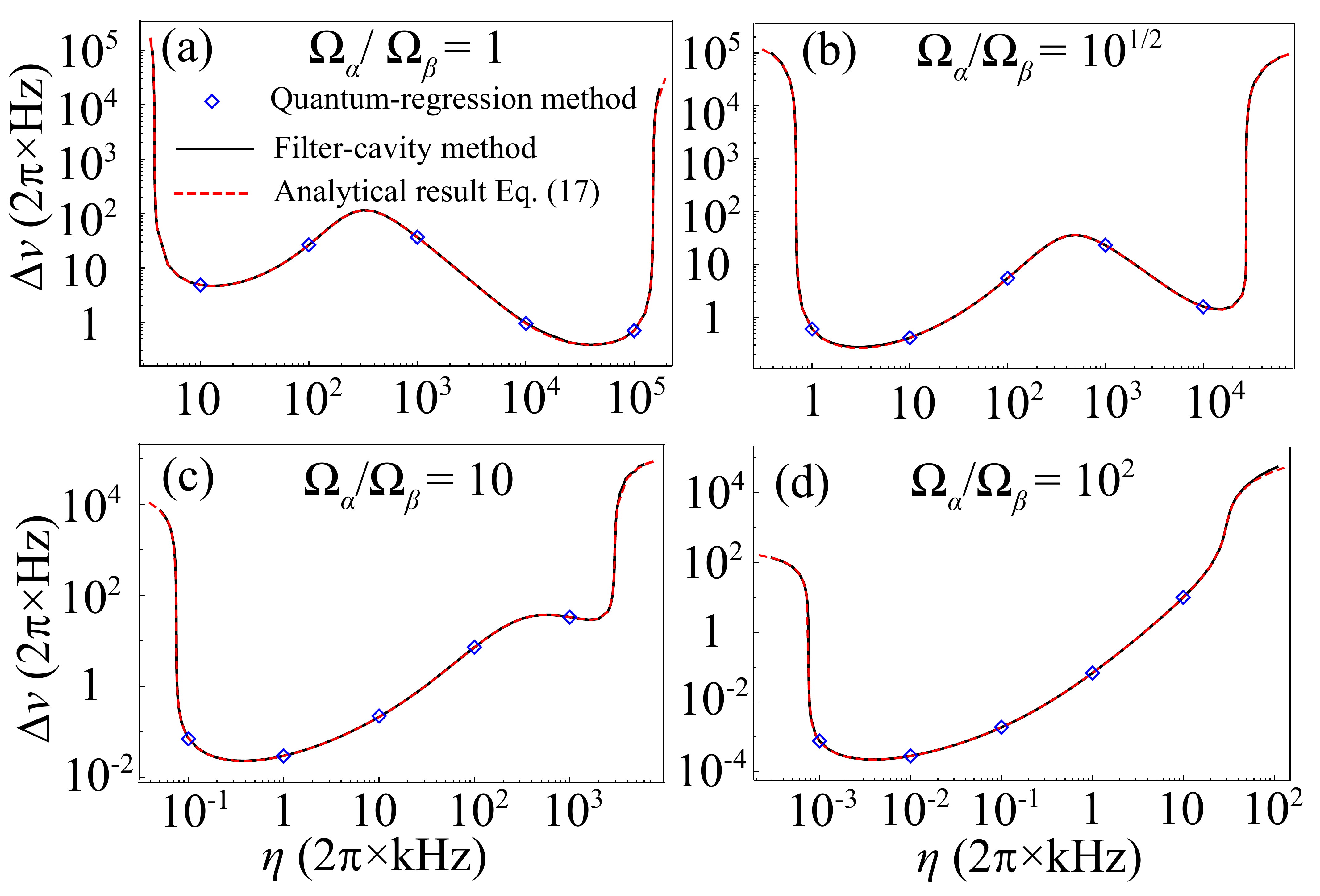}
\par\end{centering}
\caption{\label{fig:7}The laser linewidth obtained by the quantum regression
method (blue diamonds), the filter-cavity method (black lines), and the approximated analytical expression Eq.~(\ref{alw}) (red-dashed lines).
The four subplots are plotted with $\Omega_{\alpha}/\Omega_{\beta}=1$, $10^{1/2}$, $10$, and $10^2$, respectively.}
\end{figure}

\subsection{Dynamical Equations of The Correlation Functions}

The laser linewidth $\Delta\nu$ can be read from the FWHM of the laser power spectrum $S(\omega)$, which is related to the two-time correlation function of the laser according to the Wiener-Khinchin theorem~\citep{Wiener1930,Khintchine1934,Huang2009}, i.e.,
\begin{equation}
S(\omega) =2\int_{0}^{\infty}dt\textrm{Re}[\langle\hat{a}^{\dagger}(t)\hat{a}(0)\rangle_{\text{s}} e^{-i\left(\omega-\omega_{0}\right)t}].\label{eq:spectrum}
\end{equation}
Then we can use the quantum regression theorem to find the dynamical equation for the time correlation function, which reads
\begin{equation}
\frac{d}{dt}\langle\hat{a}^{\dagger}(t)\hat{a}(0)\rangle_{\text{s}} = (i\delta_{c}-\frac{\kappa}{2} )\langle\hat{a}^{\dagger}(t)\hat{a}(0)\rangle_{\text{s}}+\frac{iN\Omega_{c}}{2} \langle\hat{\sigma}_{xg}(t)\hat{a}(0)\rangle_{\text{s}}.
\end{equation}
This equation contains another correlation function $\langle\hat{\sigma}_{xg}(t)\hat{a}(0)\rangle_{\text{s}}$. Thus, we continuously derive a set of equations of motion until they are closed under the second-order mean-field approximation, as
\begin{equation}
\frac{d}{dt}A(t) =\mathbb{B}\cdot A(t),\label{eq:corr_func}
\end{equation}
where 
\begin{equation}
    A(t)=[\langle\hat{a}^{\dagger}(t)\hat{a}(0)\rangle_{\text{s}},\langle\hat{\sigma}_{xg}(t)\hat{a}(0)\rangle_{\text{s}},\langle\hat{\sigma}_{\text{P}g}(t)\hat{a}(0)\rangle_{\text{s}},\langle\hat{\sigma}_{\text{S}g}(t)\hat{a}(0)\rangle_{\text{s}}]^{\textrm{T}}
\end{equation}
and
\begin{equation}
\mathbb{B} =-\frac{1}{2}\left(\begin{array}{cccc}
-2i\delta_{c}+\kappa & -iN\Omega_{c} & 0 & 0\\
i\Omega_{c}\langle\hat{\sigma}_{xx}-\hat{\sigma}_{gg}\rangle_{\text{s}} & \gamma_{0}+\eta & 0 & -i\Omega_{\alpha}\\
i\Omega_{c}\langle\hat{\sigma}_{\text{P}x}\rangle_{\text{s}} & 0 & -2i\delta_{2}+\eta & -i\Omega_{\beta}\\
i\Omega_{c}\langle\hat{\sigma}_{\text{S}x}\rangle_{\text{s}} & -i\Omega_{\alpha} & -i\Omega_{\beta} & -2i\delta_{\alpha}+\Gamma
\end{array}\right).\\
\label{eq:B_matrix}
\end{equation}
The initial condition of the equations Eq.~(\ref{eq:corr_func}) is the steady-state solutions of the corresponding operators given in Appendix~\ref{AppA}.

\subsection{Analytical Solutions of The Dynamical Equations}

As the matrix $\mathbb{B}$ is not Hermitian, its eigenvalues are not real, and the corresponding eigenvectors are not orthogonal.
Before solving the dynamical equations Eq.~(\ref{eq:corr_func}), we need to introduce the left and right eigenvectors of $\mathbb{B}$ which are defined as
\begin{eqnarray}
\mathbb{B}\vert i\rangle &=&\lambda_{i}\vert i\rangle,\\
\langle\tilde{i}\vert \mathbb{B} &=& \lambda_{i} \langle\tilde{i}\vert. \label{eq:left_right_vec}
\end{eqnarray}

In the above equations, $\lambda_{i} (i=1,2,3,4)$ is the $i$-th eigenvalue of $\mathbb{B}$.  $\langle\tilde{i}\vert$ and
$\vert i\rangle$ are the left and right eigenvectors of $\mathbb{B}$ and satisfy the relation $\langle\tilde{i}\vert j\rangle=\delta_{i,j}$ for ($i,j=1,2,3,4$). Then the unit operator in this space becomes
$I=\sum_{i}\vert i\rangle \langle\tilde{i}\vert$.

Next, we perform the Laplace transformation to Eq.~(\ref{eq:corr_func}) and obtain
\begin{equation}
\bar{A}\left(p\right)=\frac{1}{p-\mathbb{B}}A(0),\label{eq:laplace}
\end{equation}
where $\bar{A}\left(p\right)$ is the Laplace transform of $A(t)$.

Inserting the operator $I=\sum_{i}\left|i\right\rangle \tilde{\left\langle i\right|}$ into Eq.~(\ref{eq:laplace}), we have
\begin{equation}
\bar{A}(p)=\sum_{i}\frac{1}{p-\lambda_{i}}\vert i\rangle \langle\tilde{i}\vert A(0).
\end{equation}
After applying the inverse Laplace transformation, we find the  $A(t)$ as
\begin{equation}
A(t) =\sum_{i}e^{\lambda_{i}t}\vert i\rangle \langle\tilde{i}\vert A(0).\label{eq:Atau}
\end{equation}

Eq.~(\ref{eq:Atau}) indicates that the two-time correlation function $\langle\hat{a}^{\dagger}(t)\hat{a}(0)\rangle_{\text{s}}$ is
a superposition of the functions $e^{\lambda_{i}t}\ (i=1,2,3,4)$. As the spectrum function
$S(\omega)$ is the Fourier transform of the correlation function
$\langle\hat{a}^{\dagger}(t)\hat{a}(0)\rangle_{\text{s}}$ (see Eq.~(\ref{eq:spectrum})), it is superposed by four Lorentzian lineshapes
whose central frequencies and linewidths are $\omega_{0}+\text{Im}[\lambda_{i}]$ and $2\vert\text{Re}[\lambda_{i}]\vert\ (i=1,2,3,4)$, respectively.

\subsection{Approximate Analytical Expression of Linewidth}

According to the numerical calculation, under the resonant condition that $\delta_c=\delta_\alpha=\delta_\beta=0$, the central frequency $\omega^{\ast}$ of the output laser equals the atomic $^1$S$_0$-to-$^3$P$_1$ transition frequency $\omega_0$. This fact indicates that, among the four Lorentzian lineshapes, the one corresponding to the eigenvalue with zero imaginary part and small real part contributes most significantly to the laser spectrum.

Therefore, we try to find the eigenvalue $\lambda_{\mathrm{min}}$ of the matrix $\mathbb{B}$ which has zero imaginary part and small real part.
By solving the eigenfunction of $\mathbb{B}$ to the first order, we obtain an approximated expression
\begin{equation}
\lambda_{\mathrm{min}}\simeq\frac{iA_{1}+A_{2}}{2B_{1}},    
\end{equation}
where $A_{1}$, $A_{2}$ and $B_{1}$ are real numbers, and
\begin{eqnarray*}
B_{1} &\simeq& N\Omega_{c}^{2}[\Omega_{\alpha}\mathrm{Im}\langle\hat{\sigma}_{\mathrm{S}x}\rangle_{\mathrm{s}}-(\Gamma+\eta)\langle\hat{\sigma}_{xx}-\hat{\sigma}_{gg}\rangle_{\mathrm{s}}]\\
 &&+\kappa[\eta\Gamma+(\gamma_{0}+\eta)(\Gamma+\eta)]+\gamma_{0}\Omega_{\beta}^{2}\\
 &&+(\eta+\kappa)(\Omega_{\alpha}^{2}+\Omega_{\beta}^{2})+\eta\Gamma(\gamma_{0}+\eta),\\
A_{1} &\simeq& N\Omega_{c}^{2}\Omega_{\alpha}[\Omega_{\beta}\text{Im}\langle\hat{\sigma}_{\text{P}x}\rangle_{\mathrm{s}}-\eta\text{Re}\langle\hat{\sigma}_{\text{S}x}\rangle_{\mathrm{s}}],\\
A_{2} &\simeq& \kappa[(\gamma_{0}+\eta)\Omega_{\beta}^{2}+\eta\Omega_{\alpha}^{2}+\eta\Gamma(\gamma_{0}+\eta)]\\
 &&- N\Omega_{c}^{2}(\eta\Gamma+\Omega_{\beta}^{2})\langle\hat{\sigma}_{xx}-\hat{\sigma}_{gg}\rangle_{\mathrm{s}}\\
 &&+ N\Omega_{c}^{2}\Omega_{\alpha}[\Omega_{\beta}\text{Re}\langle\hat{\sigma}_{\text{P}x}\rangle_{\mathrm{s}}+\eta\text{Im}\langle\hat{\sigma}_{\text{S}x}\rangle_{\mathrm{s}}].
\end{eqnarray*}

In the steady state, we numerically find that $\text{Re}\langle\hat{\sigma}_{\text{S}x}\rangle_{\mathrm{s}}=\text{Im}\langle\hat{\sigma}_{\mathrm{P}x}\rangle_{\mathrm{s}}=0$, which verifies the fact 
that the imaginary part of $\lambda_{\mathrm{min}}$ is zero. As the laser linewidth depends on the real part of $\lambda_{\mathrm{min}}$, we finally obtain the approximate analytical expression of the linewidth as
\begin{equation}
\Delta\nu =2\vert\text{Re}[\lambda_{\mathrm{min}}]\vert=\vert A_{2}/B_{1}\vert,\label{eq:linewidth}
\end{equation}
the explicit expression of which is given in Eq.~(\ref{alw}) of the main text.

We show the above analytical expression (all see Eq.~(\ref{alw})) with the red-dashed lines in Fig.~(\ref{fig:7}). The numerical results using the quantum-regression method and the filter-cavity method are also presented with the blue diamonds and the black-solid lines, respectively. The curves of the three methods coincide with each other very well and thus demonstrate the validity of our analytical expression.

\bibliographystyle{apsrev4-1}

\end{document}